\documentclass[aps,pra,floatfix,twocolumn,superscriptaddress]{revtex4-1}
\usepackage[english]{babel}
\usepackage{amsmath}
\usepackage{amssymb}
\usepackage{tensor}
\usepackage{graphicx}
\usepackage{color}

\newcommand{\be}{\begin{eqnarray}}
\newcommand{\ee}{\end{eqnarray}}
\newcommand{\p}{\partial}

\newcommand{\hc}{\hat{c}}
\newcommand{\dc}{\hat{c}^{\dagger}}

\begin{document}
	
\title{Mobility edges in non-Hermitian models with slowly varying quasi-periodic disorders}

\author{Qiyun Tang}
\affiliation{College of Physics, Sichuan University, Chengdu, Sichuan 610064, China}

\author{Yan He}
\affiliation{College of Physics, Sichuan University, Chengdu, Sichuan 610064, China}
\email{heyan$_$ctp@scu.edu.cn}

\begin{abstract}
We investigate the appearance of mobility edges in a one-dimensional non-Hermitian tight-banding model with alternating hopping constants and slowly varying quasi-periodic on-site potentials. Due to the presence of slowly varying exponent, the parity-time (PT) symmetry of this model is broken and its spectra is complex. It is found that the spectrum of this model can be divided into three different types of patterns depending on the magnitude of the quasi-periodic potential. As the amplitude of the potential increases from small to large, the initially well defined mobility edges become blurred gradually and then eventually disappear for large enough potential. This behavior of the mobility edges is also confirmed by a detailed study of the winding number of the complex spectra of this non-Hermitian model.
\end{abstract}

\maketitle

\section{Introduction}

The phenomena of disorders have been extensively studied in the condensed matter physics. It is well known that in three-dimensional systems, strong enough disorders will make the wave-functions of the system localized, which is the famous the Anderson localization\cite{Anderson,scaling1,scaling2}. When the disorders are not very strong, the extended and localized states coexist in the spectra and they are usually separated by the mobility edges \cite{Mott}. In the subsequent developments, people try to search for the mobility edges in low-dimensional systems but did not success. The reason is that at low dimensions even infinitesimally weak disorders can drive the whole spectra to complete localizations.

Due to these failures, people turned to one-dimensional (1D) quasi-periodic systems which possess correlated disorders. One of the paradigmatic example is the Aubry-Andre (AA) model \cite{AA,Harper}, which is a 1D hopping model with incommensurate on-site potentials. One important feature of the AA model is the existence of the self-duality at certain disorder strength. When the disorder potential increases beyond this point, all the extended eigenstates suddenly turns into localized ones. Although the mobility edges are absent in the original AA model, they do exist in some generalized AA models with long range hopping or unbounded potentials \cite{Biddle,Ganeshan,Izrailev,Biddle-11}. In these models, The self-duality can also be employed to determine the exact shape of mobility edges. Later on, a large class of models with slowly varying quasi-periodic disorders \cite{Xie1,Xie2,Li-17,Deng-19,Saha-19} was introduced to support mobility edges. But the self-duality is broken in this type of models.

The recent years witness a rapid developments of non-Hermitian physics, which has been applied to almost all aspects of condenses matter physics \cite{Gong,Kawabata}. The non-Hermitian systems can display dramatically different properties in both topology and symmetry \cite{Esaki,TE-Lee,HT-Shen,Wang,Ghatak_2019,Jiang-2019}. The early work that generalized disorder systems to the non-Hermitian case is the Hatano-Nelson model\cite{Hatano1,Hatano2}. After that, there appeared a lot of works that has generalized the original AA model to non-Hermitian lattices \cite{Longhi-2019,ChenShu-17,ChenShu-20,Liu-22,Yuce,Liu-20,CX-21,TangLZ-21,Amin-01,Xu-21}. In most of these non-Hermitian quasi-periodic models, it is found that the extended or localized states usually correspond to the real or complex spectra of the systems \cite{YanXia-21,Xia-22,LiuYanxia-21,ChenWen-22,Wang-22,GUO-23}. In the same time, the self-duality is an important tool to determine the mobility edges \cite{Liu-20,Gopalakrishnan}.

Although the non-Hermitian quasi-periodic models are widely investigated, the study of their slowly varying counterpart is relatively rare. In this paper, we present a detailed study on the non-Hermitian Su-Schrieffer-Heeger (NH-SSH) model \cite{ChenShu-SSH} with slowly varying quasi-periodic on-site potentials. Because of the slowly varying potential, the parity-time (PT) symmetry is explicitly broken and the spectrum of this model is always complex. The property of self-duality which is a signature of localization transition is also lost in this model.

Despite of the failure of self-duality, we can still semi-analytically determine the mobility edges by the so-called ``energy matching method'', which we have applied to Hermitian slowly vary models in our previous works \cite{Tang-21,Tang-23}. Intuitively speaking, this method approximate the quasi-periodic model by a set of different periodic models. Then the region of extended states of the quasi-periodic model can be obtained by taking intersection of the energy bands of all these periodic models. According to the spectra of these periodic models, we find that the mobility edges displays 3 types of behaviors: well defined, blurred or completely disappeared. Since the spectrum of the above NH-SSH model is complex, we can also compute the winding number of this complex spectrum. The dependence of the winding number on the disorder potential also displays 3 different trends, which exactly matches the 3 types of behaviors of mobility edges we mentioned before.

The rest of this paper is organized as follows. In section \ref{sec-SSH}, we introduce the non-Hermitian SSH model with slowly varying quasi-periodic disorders, which is the main focus of this paper. In section \ref{sec-num}, we will first employ the energy matching method to qualitatively understand the spectra and mobility edges of our slowly varying quasi-periodic NH-SSH model. Then the energy spectra and Inverse Participate Ratio (IPR) indices are numerically calculated for this model in order to confirm the results of previous semi-analytical analysis. In section \ref{sec-wind}, we make use of the winding number as another index to understand the behaviors of mobility edges from another angle. In the end, we briefly conclude in section \ref{sec-conclu}.

\section{The Slowly Varying Quasi-Periodic Non-Hermitian SSH model}
\label{sec-SSH}

In this section, we define the non-Hermitian SSH model with slowly varying quasi-periodic complex on-site potentials. We will focus on the energy spectrum and the behaviors of mobility edges of this model by varying its parameters. Later, we will also briefly discuss some possible generalization of this type of model.  The Hamiltonian of this model can be written as
\be
\hat{H}&=&\sum_{n=1}^{L/2}\Big[(t_1\dc_{n,A} \hc_{n,B}+h.c.)+V_{n,A}\dc_{n,A} \hc_{n,A}\nonumber\\
&&+V_{n,B}\dc_{n,B} \hc_{n,B}\Big]+\sum_{n=1}^{L/2-1}\Big(t_2\dc_{n,B} \hc_{n+1,A}+h.c.\Big)
\label{SSH1}
\ee
Here $\dc_{n,A/B}(\hc_{n,A/B})$ is the fermion creation (annihilation) operator at site $n$ and orbital $A$ or $B$. $L$ is the total number of lattices sites in the model. $t_1=t-\lambda$ and $t_2=t+\lambda$ are the intra-cell and inter-cell hopping constants respectively. Here $V_{n,A}$ and $V_{n,B}$ are the on-site complex quasi-periodic potential which is given by
\be
V_{n,A}=Ve^{(i2\pi\alpha n^v)},\quad V_{n,B}=Ve^{(-i2\pi\alpha n^v)}
\label{SSH2}
\ee	
Note that $V_{n,B}$ is the complex conjugate of $V_{n,A}$.  Here $\alpha$ is an irrational number, which introduces quasi-periodic disorders. Throughout the whole paper, we assume that $\alpha=(\sqrt{5}-1)/2$. In the same time, we also introduced an slowly varying exponent $v$ satisfying $0<v<1$. The introducing of slowly varying exponent breaks the PT symmetry. If $V_{n,A}=V_{n,B}=0$, then the model gets back to the standard Hermitian SSH model. If we set $\lambda=0$, Eq.(\ref{SSH1}) becomes a  non-Hermitian AA model with slowly varying quasi-periodic disorders. This special case will also be studied in later sections.

In order to achieve slowly varying, we always assume that $0<v<1$. With this condition, it is easy to see that the derivatives of the potential $V_{n,A}$ approach to zero as the site index become very large.
\be
\lim_{n\to\infty}\frac{d V_{n,A}}{d n}=\lim_{n\to\infty}i 2\pi \alpha v n^{v-1} V e^{(i2\pi\alpha n^v)}=0
\ee
Similar result is also valid for $V_{n,B}$. This indicates that the quasi-periodic potential approaches a constant when the number of lattice sites of the model become very large. But for a given finite sized model, this constant is not unique and varies within certain range depending on the system size. Because of this behavior, we claim that this model contains slowly varying quasi-periodic disorders. The existence of this slowly varying quasi-periodic disorders makes the appearance of the mobility edge possible in our model even without PT symmetry.

\section{mobility edges of the non-Hermitian SSH model}
\label{sec-num}

\subsection{Determine the mobility edges by energy matching method}

In this subsection, we will present a heuristic method called ``energy matching method''\cite{Tang-21}, which can efficiently  determine the mobility edges in Hermitian slowly varying quasi-periodic models. With some refinement, we found that this method is also applicable for the NH-SSH model. In addition, it can provide some clue for what type of slowly varying quasi-periodic models can support mobility edges when generalized to non-Hermitian systems. The core idea of this method is to make use the characteristics of slowly varying quasi-periodic potential that approaches to a constant at very large site index. Because of this, we can approximate the slowly varying quasi-periodic disordered model by a set of periodic models. The approximated periodic model Hamiltonian can be expressed as follows
\be
\hat{H}&=&\sum_{n=1}^{L/2}\Big[(t_1\dc_{n,A} \hc_{n,B}+h.c.)+V_M^*\dc_{n,A} \hc_{n,A}\nonumber\\
&&+V_M\dc_{n,B} \hc_{n,B}\Big]+\sum_{n=1}^{L/2-1}\Big(t_2\dc_{n,B} \hc_{n+1,A}+h.c.\Big)
\label{eq-per}
\ee
Here $V_M=Ve^{(i2\pi \alpha M^v)}$ and $V_M^*$ is the complex conjugate of $V_M$. For a fixed $M$, the potential is constant, therefore $\hat{H}_M$ describes a periodic model. To reflect the quasi-periodic disorders of original model, we allow $M$ to take all integer values between $1$ and $L$. For convenience, we set $2\pi \alpha M^v=\phi$. It is obvious that the above periodic Hamiltonian has no non-Hermitian skin effect. Therefore all the eigenstates of the above periodic models are extended.

For these periodic model of Eq.(\ref{eq-per}), we can diagonalize the Hamiltonian by transferring to the momentum space, and find out that energy bands are given by
\be
E=V\cos \phi\pm\sqrt{t_1^2+t_2^2+2t_1t_2\cos k-V^2\sin^2 \phi}
\label{band}
\ee
For a given periodic model, the eigenstates with the energy inside the band are all extended. Here, we will focus on the intersection of the energy bands of all the periodic models. As we discussed before, the quasi-periodic model can be thought as a combination of all these periodic modes in some sense. Therefore, if the eigen-energy of the quasi-periodic model fall within the range of the above intersection, we can expect that the corresponding eigenstates to be extended. The region of extended states is obtained by matching their eigen-energy to the bands of periodic models. Therefore, it is dubbed as the ``energy-matching method". Since $|\cos k|<1$, the energy region of the extended state is given by the following intersection
\be
&&E\in \bigcap_a\Big(V \cos\phi+\sqrt{(2\lambda)^2-V^2\sin^2\phi},\nonumber\\
&&\qquad V\cos\phi+\sqrt{(2t)^2-V^2 \sin^2 \phi}\Big)
\label{inter}
\ee
Here we only discuss the extended states region from the plus sign of Eq.(\ref{band}). For the minus sign, the discussion is very similar. For Hermitian systems, the energy band of periodic models are real, thus the intersection of energy bands can be easily obtained. In contrast, for non-Hermitian systems, the energy bands of these periodic models can be complex, so the energy-matching method needs to be further refined before it can be applied to non-Hermitian quasi-periodic models.	

We find that the energy bands of Eq. (\ref{band}) can be divided into three different classes depending on the relative magnitude of $V$ comparing to $2t$ and $2\lambda$. Accordingly, the behavior of mobility edges of the NH-SSH model of Eq.(\ref{SSH1}) also falls into 3 classes. The details are explained as follows.

(1) When $V \leq 2 \lambda$, no matter what value $\phi$ takes, the PT symmetry of the periodic model is not broken, and the energy bands of the periodic models are all real numbers. At this point, we can directly find the intersection of the energy bands of different $\hat{H}_M$, and the mobility edge is given by
\be
E\in \big(2\lambda+V, \, 2 t-V\big)
\label{eq-E}
\ee

(2) When $2\lambda<V<2t$, if $\phi$ satisfies $ \phi>|\arcsin(2\lambda/V)|$, the PT symmetry of the periodic model is broken, and the energy spectra of the periodic models have both real and complex values. Since $V<2t$, we can see that the upper bounds of each energy bands in Eq.(\ref{inter}) are still real numbers. On the other hand, since $V>2\lambda$, the lower bounds of each energy bands in Eq.(\ref{inter}) can become complex numbers. Since there is no natural orders for complex numbers,  one can not determine the precise intersection of the lower bounds in Eq.(\ref{inter}). Instead, we can use the real parts of these complex eigen-energy to determine the lower bound as follows
\be
&&E\in \bigcap_a\Big(\mbox{Re}\Big[V \cos\phi+\sqrt{(2\lambda)^2-V^2\sin^2\phi}\Big],\nonumber\\
&&\qquad V\cos\phi+\sqrt{(2t)^2-V^2 \sin^2 \phi}\Big)
\ee
Then it is easy to find the region of extended states as
\be
E\in \big(2\lambda+V, \, 2 t-V\big)
\ee
The mobility edge obtained in this way seems to be the same as Eq.(\ref{eq-E}) of the type (1). But the mobility edge in this case is actually not completely correct, because we only find the intersection of the real part of the complex eigen-energys and ignore the effects of the imaginary parts. We will demonstrate this in-accuracy by the numerical calculations in the next subsection.

(3) When $V \geq 2t$, if $\phi$ satisfies $\phi>|\arcsin(2 t/V)|$, the PT symmetry of the model is completely broken and all eigen-energy of the periodic models are complex. In addition, if we look for the intersection of the real parts of these complex energy bands, the result is an empty set. In other words, there is no extend states any more. Therefore, mobility edges are no longer existed in this case.

In the next subsection, the numerical calculations will be carried out to verify the mobility edges that are obtained by the above energy-matching method. These results also demonstrate that it is reasonable to divide the behaviors of mobility edges into three different classes.

\begin{figure*}[t]
\centering
\includegraphics[width=\textwidth]{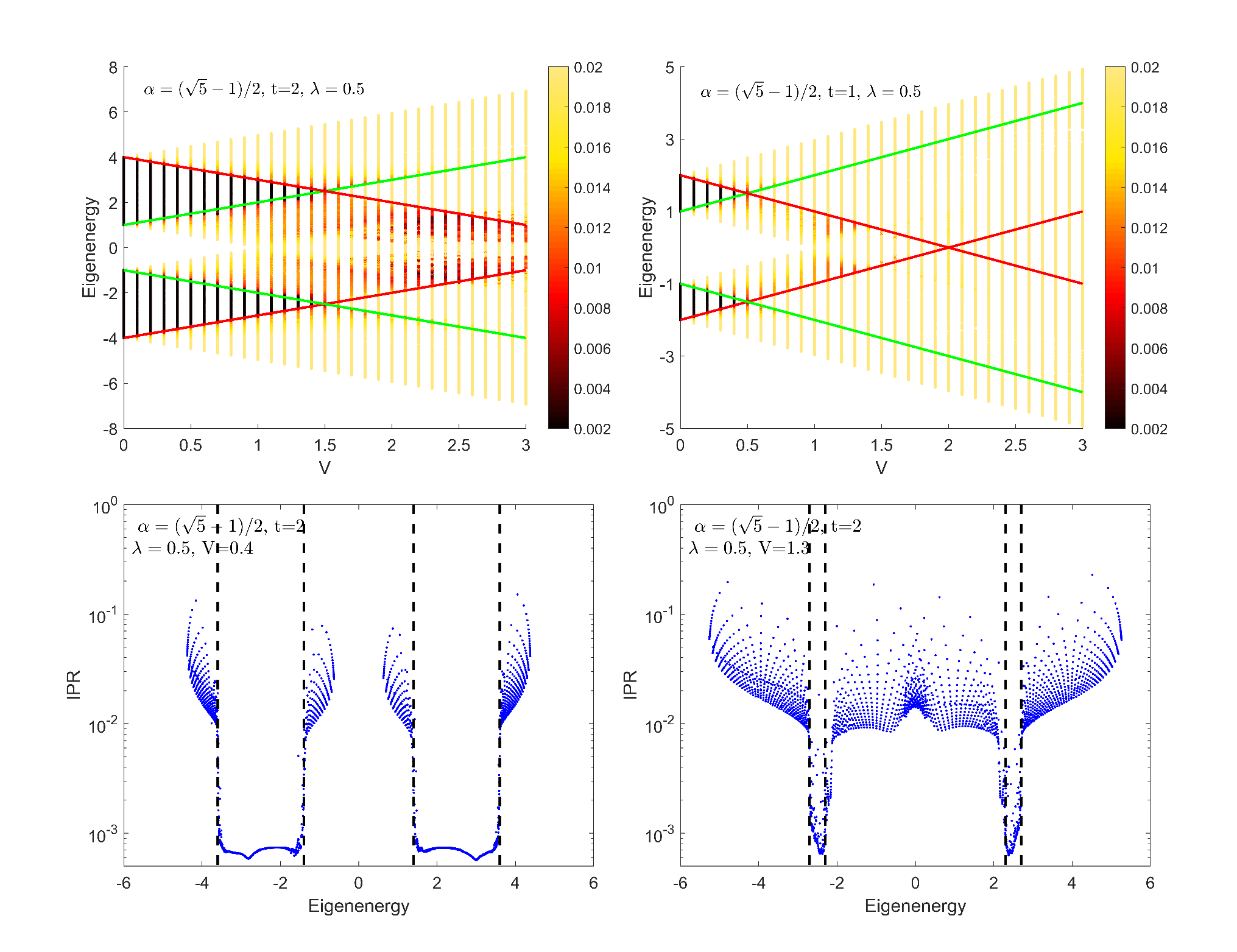}
\caption{Top row: The real parts of the energy band of Eq.({\ref{SSH1}}) as a function of $V$. We assume that $t=2$ (top left) and $t=1$ (top right) respectively. The color of the point represents its IPR value. The red and green lines represents the mobility edges. Bottom row: the IPR as a function of the eigenenergy. We assume $V=0.4$ (bottom left) and $V=1.3$ (bottom right) respectively. The dotted lines indicate the mobility edges. Other parameters used in calculations are the $\alpha=(\sqrt{5}-1)/2$, $v=0.5$, $\lambda=0.5$ and $L=2000$.}
\label{fig-band}
\end{figure*}

\begin{figure*}[t]
\centering
\includegraphics[width=\textwidth]{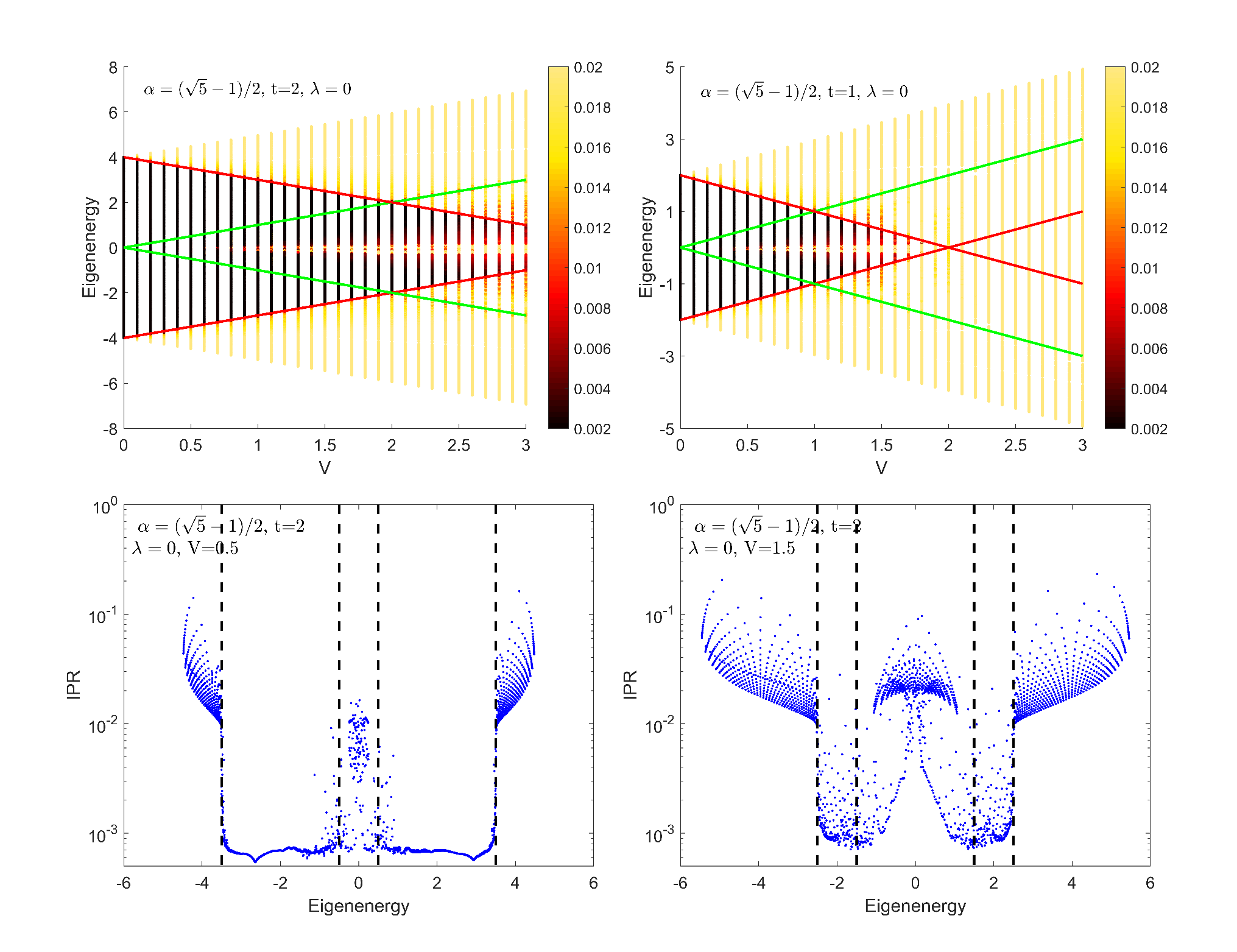}
\caption{Top row: The real parts of the energy band of Eq.({\ref{SSH1}}) as a function of $V$ with $\lambda=0$. We assume that $t=2$ (top left) and $t=1$ (top right) respectively. The color of the point represents its IPR value. The red and green lines represents the mobility edges. Bottom row: the IPR as a function of the eigenenergy. We assume $V=0.5$ (bottom left) and $V=1.5$ (bottom right) respectively. The dotted lines indicate the mobility edges. Other parameters are the same as in Fig.\ref{fig-band}}
\label{fig-lam0}
\end{figure*}

\subsection{Numerical results of the mobility edges}

In order to demonstrate to what degree the eigenstates of the model localized, we make use of the Inverse Participate Ratio (IPR) as the indicator of the localization \cite{Thouless,Kohmoto}. For the $m$-th normalized wave function $\psi_m$, the IPR is defined as follows
\be
\mbox{IPR}_m=\sum_{j=1}^L\Big|a^m_j\Big|^4,\quad \psi_m=(a^m_1,\cdots,a^m_L)
\ee
Here $L$ is the system size. It is well known that, for a generic extended state, the amplitude at each site should be roughly uniform. Therefore, if the eigenstate is extended, its IPR should be close to
\be
\sum_{j=1}^{L}(|a_j^m|^2)^2\sim\sum_{j=1}^{L}\frac{1}{L^2}=\frac{1}{L}\nonumber
\ee
which is close to $0$ as $L\to \infty$. For a generic localized state, on the other hand, the amplitude will concentrate around a few particular sites. Therefore, one expected that the IPR of a localized state should be a number with order one.

We plot the real part of the energy band of of Eq.{\ref{SSH1}} as a function of $V$ in the top row of Fig. $\ref{fig-band}$. The color of each point represent the IPR value of the corresponding eigenstate. The parameters used in the calculations are listed in the figure caption. The red and green lines are the mobility edges of Eq.(\ref{eq-E}) obtained by the energy matching method. It can be seen that these lines agree well with the separation boundary between dark colored and light colored areas in the figure. The dark and light colors correspond to the extended and the localized state regions respectively.

It is useful to make a closer examination of the transition between the extended state and the localized state near the mobility edge. This can clearly demonstrate the differences among the 3 behaviors of the mobility edges discussed in the previous subsection. We plot the IPR as a function of the eigenenergy for two selected $V$ in the bottom row of Fig. $\ref{fig-band}$. In the bottom left panel, we choose $V=0.4$ such that $V<2\lambda$. In the bottom right panel, $V=1.3$ satisfies $2\lambda<V<2t$. From the bottom row of Fig. \ref{fig-band}, we can clearly see that there is a significant jump of roughly $10^{-2}$ in the IPR values near the mobility edge, which shows the transition between the extended state and the localized state. It confirms the mobility edges in the pseudo-color plots in the top row of Fig. \ref{fig-band}. In addition, one can see that there are some points in the extended state energy region that are significantly different from the IPR value of the extended state by the orders of magnitude in the bottom right panel of Fig.$\ref{fig-band}$. These points are actually localized states. The existence of localized states inside the extended state region indicates that the mobility edge becomes blurred, which is the signature of the second class that we have discussed before.

We would like to emphasise that the second class where $2\lambda<V<2t$ is special for the slowly varying quasi-periodic non-Hermitian models.
In order to take a closer look at this region, we set $\lambda=0$, then the 3 classes of behaviors separated by $2\lambda$ and $2t$ collapses into two classes. First, when $V<2t$, the eigen-energys of some periodic models in the energy-matching method are both real and complex. Second, for $V>2t$, the eigen-energys of some periodic models are all complex numbers. We plot the real part of the energy band of Eq.{\ref{SSH1}} as a function of $V$ for $t=2$ (left) and $t=1$ (right) in the top row of Fig. $\ref{fig-lam0}$. Again, The color of the point represents its IPR value. One can see that the mobility edge obtained by the energy matching method is not completely correct. More specifically, the mobility edge represented by the red lines correctly separate the localized states from the extended states. But the green lines fail to do so. This is consistent with our previous analysis that the lower bound of the extended states region does not exist due to the appearance of complex numbers eigenvalues. These complex eigenvalues invalidates the mobility edge obtained from the intersection of the lower bounds of Eq.(\ref{inter}).

Similarly, we plot the IPR as a function of the eigen-energy for two selected $V$ in bottom row of Fig. $\ref{fig-lam0}$. One can see that there is an clear transition between the extended states and the localized states near the red line (represented by the two outer dashed lines). But the IPR values near the green line (represented by the two inner dashed lines) are quite chaotic. Therefore, the mobility edges corresponding to the green line is no longer valid, which is consistent with our previous analysis. Moreover, as $V$ increases, the number of periodic models with complex eigenvalues also increases. This suggests that the mobility edges will become even more blurred as $V$ increases. This behavior is also reflected in bottom row of Fig. $\ref{fig-lam0}$.

\section{winding numbers of the non-Hermitian SSH model}
\label{sec-wind}

\begin{figure}[t]
\centering
\includegraphics[width=0.7\columnwidth]{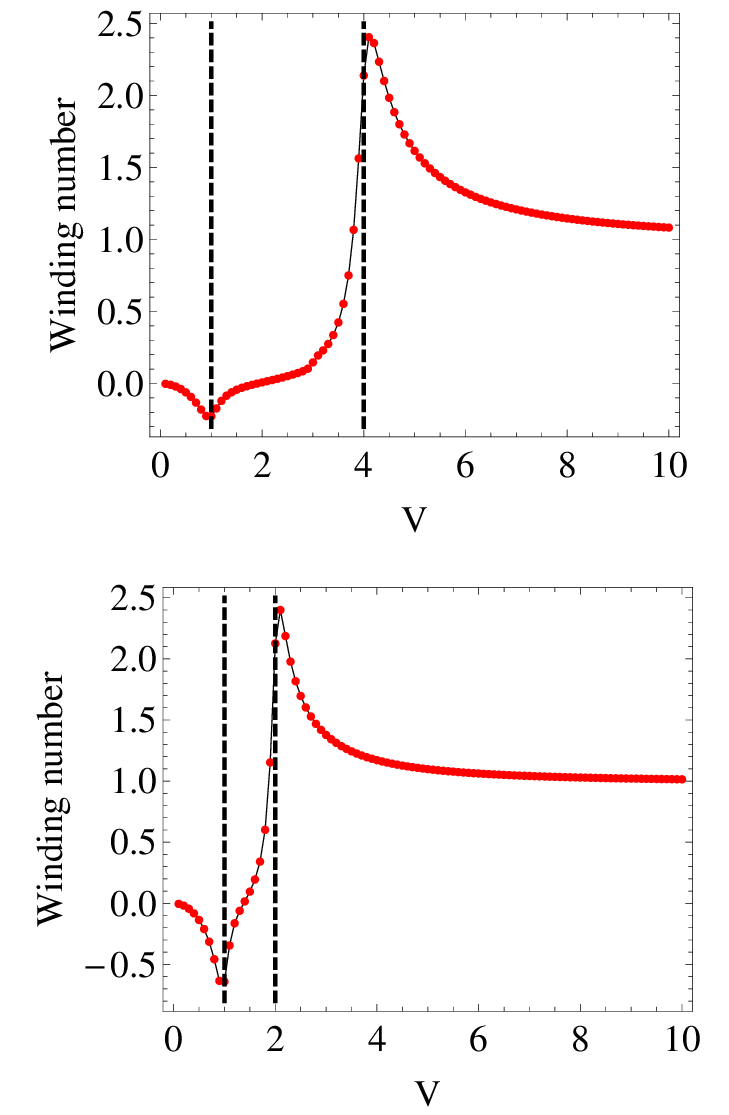}
\caption{The winding number for the model of Eq. (\ref{SSH1}) as a function of $V$  with $t=2$ (top) and $t=1$ (bottom), respectively. Other parameters are $\lambda=0.5$, $\alpha=(\sqrt{5}-1)/2$, $v=0.5$ and $L=600$. The dotted lines are $V=2\lambda$ and $V=2t$.}
\label{wind-1}
\end{figure}

\begin{figure}[t]
\centering
\includegraphics[width=0.7\columnwidth]{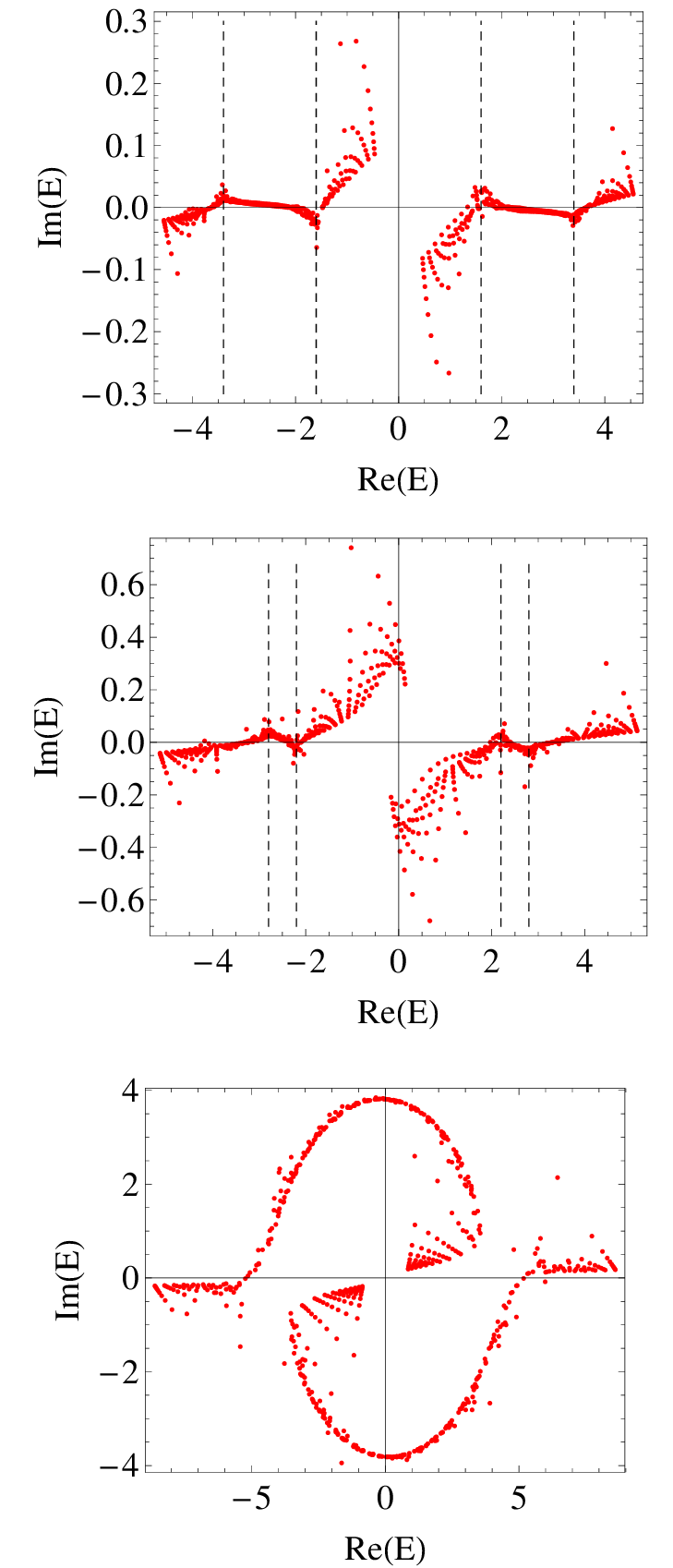}
\caption{The eigen-energy $E$ of Eq. (\ref{SSH1}) plotted on the complex plane. From top left panel to the bottom right panel, $V=0.6$, $1.2$ and $4.8$, respectively. The dashed lines represents the mobility edges. The parameters are the same as the top panel of Fig.\ref{wind-1}.}
\label{ev-1}
\end{figure}

For quasi-periodic non-Hermitian systems with PT symmetry, the winding number is considered as an efficient tool to characterize the localization of the system. If the system is extended, the winding numbers is trivial $w=0$. One the other hand, if the system is localized, it possess non-trivial winding numbers such as $W=\pm1$. Therefore, we could also calculate the winding numbers for the model of Eq. (\ref{SSH1}) to investigate whether the model is localized or extended \cite{Xu-20,Han-22,Cai-21,Cai-22,zhou-23,Longhi-19}. In order to define the winding number, we have to introduce an additional phase parameter $\theta$ into the potential of the model. More specifically, we can make the following replacement in Eq.(\ref{SSH1})
\be
V_{n,A}\to Ve^{(i2\pi\alpha n^v+i\frac{\theta}{L})},\quad
V_{n,B}\to Ve^{(-i2\pi\alpha n^v+i\frac{\theta}{L})}\nonumber
\ee
Then we arrived at a Hamiltonian that depends on $\theta$, which is denoted as $H(\theta /L)$.
For a selected reference energy $E_b$, the winding numbers is defined as \cite{Jiang-2019,Gong}
\be
w(H)=\lim_{L \to \infty} \int_{0}^{2\pi} d\theta \frac{\p}{\p\theta}\log\det \Big[H(\theta /L)-E_b\Big]
\label{eq-wind}	
\ee	
In our calculation, we set $E_b=0$.

The intuitive picture of the winding number is to count how many times the complex energy spectrum of the model circles around the reference energy $E_b$, as $\theta$ increases from 0 to $2\pi$. Usually, the winding number can serve as an indicator to the extended-localized transitions in non-Hermite systems with PT symmetry. This is because that the localization transition in non-Hermitian model is often accompanied by the PT symmetry breaking. Then the spectrum of the non-Hermitian model become complex and tend to exhibit a circular structure in the complex plane. Therefore, By observing whether the spectrum of the model exhibits a circular structure through the winding number, one can efficiently determine whether the system becomes localized. However, in our NH-SSH model, the PT symmetry is explicitly broken by the slowly varying potentials. We will see that the winding number exhibits some new phenomena in our model.

In Fig. \ref{wind-1}, we plot the winding number as a function of $V$ for $t=2$ (top) and $t=1$ (bottom), respectively. It is evident that there are two singular points located at $V=2\lambda$ and $V=2t$ corresponding to the two dashed lines in the figure. One can see that the two singular points divide $V$ axes into three parts, which agrees with the three types of behaviors of mobility edges we discussed before. Here we can again use the energy matching method to qualitatively understand the behavior of the winding number. The advantage of the approximation of Eq.(\ref{SSH1}) by a set of periodic models is that we have simple expression Eq.(\ref{band}) for the spectra of the periodic models.

For $V<2\lambda$, the spectra of the approximated periodic models are all real. Therefore, in this case, the winding numbers is almost 0. Meanwhile, the intersection of all these bands give rise to the region of extended states. When $V$ is close to the first singular point at $2\lambda$, the winding number deviates from 0, which suggest some qualitative changes happen to the system around this point. When $2\lambda<V<2t$, the spectra of the periodic models contains both real and complex eigenvalues. Accordingly, the winding numbers is not stable and rapidly increases from a very small value to a very large value. Meanwhile, the mobility edges become blurred in this case. When $V>2t$, the spectra of the periodic models are all complex. Now the winding number decreases from the extreme large value and finally stabilizes to the quantized value 1. In the same time, the mobility edges disappear in this case, which means that all eigen-state are localized now. Therefore, the above 3 cases are consistent with the 3 types of behaviors of mobility edges.

\begin{figure}[t]
\centering
\includegraphics[width=0.7\columnwidth]{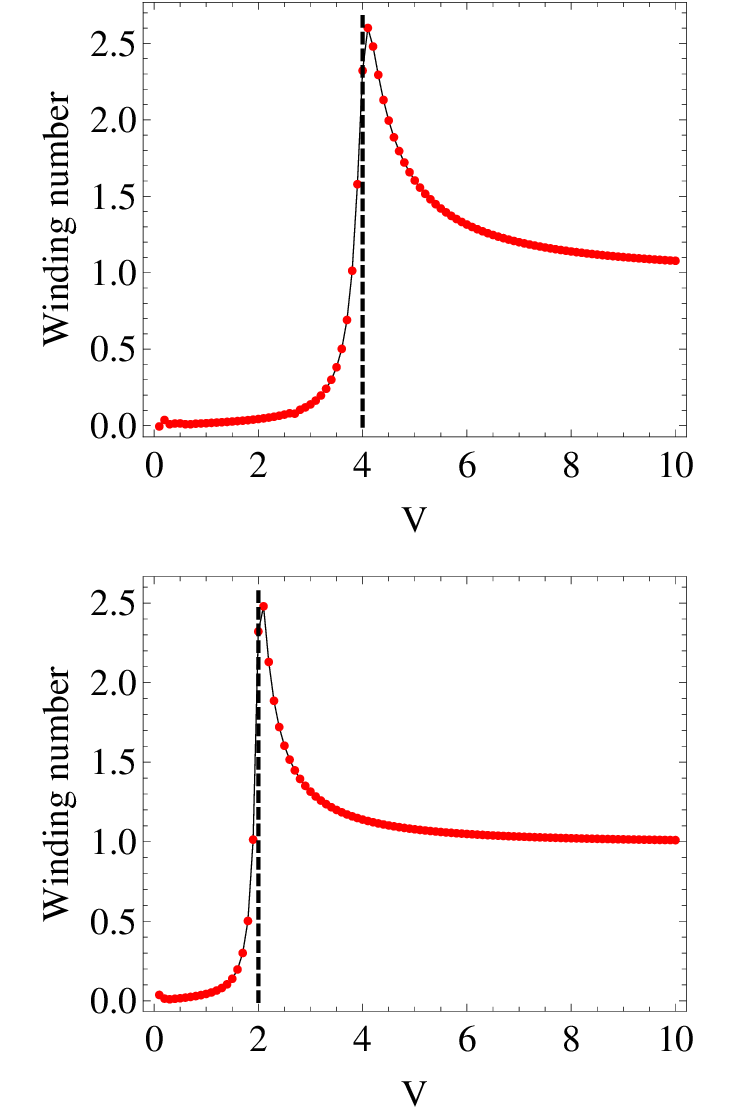}
\caption{The winding number for the model of Eq. (\ref{SSH1}) as a function of $V$  with $t=2$ (top) and $t=1$ (bottom), respectively. Other parameters are $\lambda=0$, $\alpha=(\sqrt{5}-1)/2$, $v=0.5$ and $L=300$. The dotted lines are $V=2\lambda$ and $V=2t$.}
\label{wind-2}
\end{figure}

\begin{figure}[t]
\centering
\includegraphics[width=0.7\columnwidth]{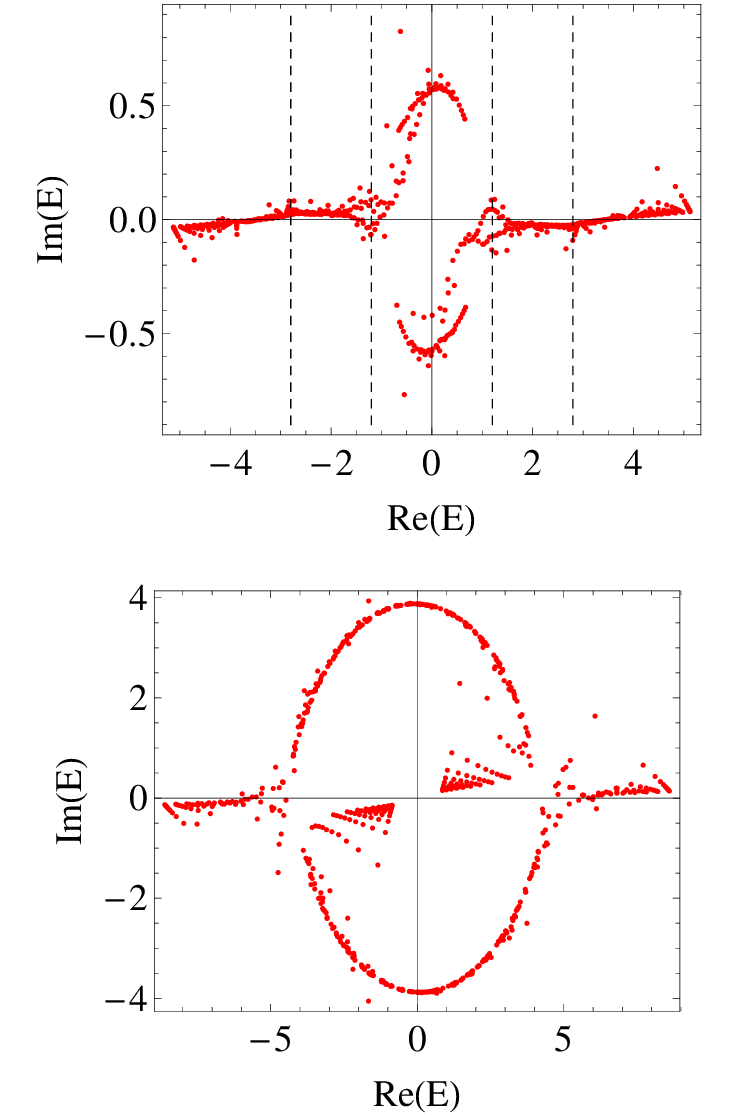}
\caption{The eigen-energy $E$ of Eq. (\ref{SSH1}) plotted on the complex plane. From top left panel to the bottom right panel, $V=1.2$ and $4.8$, respectively. The dashed lines represents the mobility edges. The parameters are the same as the top panel of Fig.\ref{wind-2}}
\label{ev-2}
\end{figure}

We can also directly investigate how the spectra of Eq.(\ref{SSH1}) evolve with $V$. In Fig. \ref{ev-1}, we plot the spectrum of Eq.(\ref{SSH1}) on the complex plane for $V=0.6$, $1.2$ and $4.8$ respectively. Due to the PT symmetry breaking of this model, the spectra are all complex. On the top panel, the dashed lines indicate the mobility edges. Although the eigen-energy of the extended states are also complex, they are very close to the real axes. On the other hand, the spectra of the localized states are far more scattered. But they are far from making a whole circle on the complex plane yet. Thus, the winding numbers is almost completely 0 for $V<2\lambda$. As $V$ increases to the range of $2\lambda<V<2t$, the number of extended states decreases and the localized states start to dominate the spectra. But their spectra still cannot make a full circle.  As $V$ further increases to $V>2t$, the spectra of the localized states finally complete a whole circle, and the winding numbers is eventually reached the quantized value 1.

We can also consider a special case by setting $\lambda=0$. Then there is only one singularity point for the windings number curve, as shown in Fig. \ref{wind-2} . This is consistent with previous analysis with $\lambda=0$, where the behavior of mobility edges also fall into two cases: $V<2t$ and $V>2t$. In Fig.\ref{ev-2}  we plot the spectrum of Eq.(\ref{SSH1}) with $\lambda=0$ on the complex plane for $V=1.2$ (top) and $4.8$ (bottom). The spectra of $V=4.8$ case is almost the same as in the bottom panel of Fig.\ref{ev-1}. However, the spectra of $V=1.2$ case of is far more scattered comparing to the middle panel of Fig.\ref{ev-1}. Due to $\lambda=0$, the spectra already contain both real and complex eigenvalues in the range of $V<2t$. Therefore, there are more localized states in this case and give rise to more scattered spectra.

In summary, we have found an intuitive relation between the mobility edges and the shape of spectra on the complex plane. If the spectra is roughly confined around the real axes, the extended states dominate the spectra and there exist sharp mobility edges. If the spectra is scattered, then the system is comprised by a mixture of both extended and localized states and the mobility edges are blurred. If the spectra almost form a circle around a fixed point, then the localized states dominate the system and there is no mobility edges.

\section{conclusion}
\label{sec-conclu}

In this paper, we present an example of non-Hermitian model with slowly varying quasi-periodic disorders which can support well defined mobility edges.  We found that the behavior of mobility edges fall into 3 classes depending on the amplitude of disorder potentials. These 3 classes also match with the winding number of the complex spectrum of the system. In order to understand these behaviors, we found that it is very useful to approximate the slowly varying model by a set of periodic models. This method is not only effective in determining the mobility edges in our example, it can also suggest a way to construct other slowly varying models to support mobility edges.

\begin{acknowledgements}
This work was supported by the Natural Science Foundation  of  China  under  Grant  No. 11874272.
\end{acknowledgements}


\end{document}